\documentstyle[mprocl]{article}
\def\JOURNAL#1#2#3#4{{#1} {\bf #2}, #3 (#4)}
\def\PRE{{\em Phys.\ Rev.}\ E}

\def\JPI{{\em J.\ Phys.\ I (France)}}
\def\JPA{{\em J.\ Phys.\ A: Math\ Gen.}}

\def\be{\begin{equation}}
\def\ee{\end{equation}}
\def\bea{\begin{eqnarray}}
\def\eea{\end{eqnarray}}

\def\vm{v_{max}}
\begin{document}
\title{CELLULAR AUTOMATA FOR TRAFFIC FLOW:\\ ANALYTICAL RESULTS}
\author{ ANDREAS SCHADSCHNEIDER }
\address{Institut f\"ur Theoretische Physik, Universit\"at zu K\"oln,\\
         50937 K\"oln, Germany}
\author{ MICHAEL SCHRECKENBERG }
\address{Theoretische Physik/FB 10, Gerhard-Mercator-Universit\"at Duisburg,\\
         47048 Duisburg, Germany}
\maketitle
\abstracts{We use analytical methods to investigate cellular automata
for traffic flow. Two different mean-field approaches are presented, which
we call site-oriented and car-oriented, respectively.
The car-oriented mean-field theory yields the exact fundamental diagram
for the model with maximum velocity $\vm =1$ whereas in the site-oriented
approach one has to take into account correlations between
nearest-neighbour sites. Going beyond mean-field using the so-called
$n$-cluster approach our results for $\vm >1$ are in excellent
agreement with numerical simulations. We also present a modified
cellular automaton which is closely related to a two-dimensional dimer
model.}

\section{Introduction}

In 1992 Nagel and Schreckenberg \cite{ns} introduced a cellular automaton
model describing traffic flow of $N$ cars on a single lane. In this model
the street is divided into $L$ boxes ('cells') of a certain length (for
realistic applications 7.5 meters) which can be occupied by at most
one car or be empty. The cars have an internal parameter ('velocity')
which can take on only integer values $v=0,1,2,\ldots,\vm$. The
dynamics of the model is described by update rules for the velocities
and the motion of the cars. The update rules
are given by the following four steps \cite{ns}:\\[0.2cm]
1) Acceleration: $v_i\rightarrow v'_i=v_i+1$ (if $v_i < \vm$)\\
2) Slowing down: $v_i\rightarrow v'_i=d_i-1$ (if $d_i \le v_i$)\\
3) Randomization: $v'_i\rightarrow v''_i=v'_i-1$ (with probability $p$)
for $v'_i>0$\\
4) Car motion: Car moves $v''_i$ sites\\[0.2cm]
Here $v_i$ is the velocity of car $i$ and $d_i$ is the distance between
cars $i$ and $i+1$. Without step 3) the dynamics would be purely
deterministic and the system shows strong dependence on the initial
condition. The randomization takes into account natural
fluctuations in the driver's behaviour.
All cars are updated simultaneously (parallel update)
and we here use periodic boundary conditions (``Indianapolis situation'').
For the analytical calculation it is usually advantagous to change the
order of the steps into 2-3-4-1, since after step 1) there are no
vehicles with velocity 0. This change in order than has to be taken
into account in the calculation of the flux (fundamental diagram).

Most of the results for this cellular automaton have been obtained from
simulations \cite{nana}. Basically there are two different methods to
implement the rules \cite{ssni}. In the car-oriented approach the state
is characterized by the occupancy of each individual site (which might
be empty or occupied by a car with velocity $0,1,\ldots,\vm$). In the
car-oriented method on the other hand one has two lists, one with the
velocities of the cars and one in which the distance to the next car
ahead is stored. For small densities $c=N/L$ the last method is preferable
in simulations and for higher densities the first one.

\section{Site-Oriented Approach}

\begin{figure}
\vskip 60mm
\includegraphics{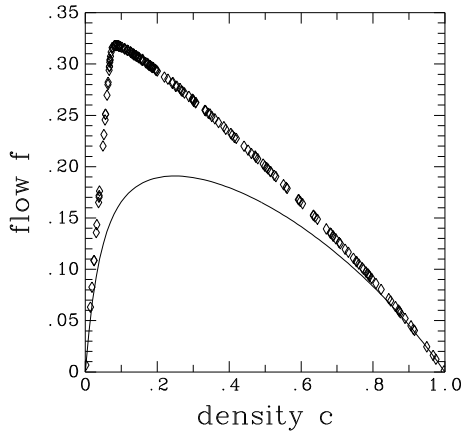}
\caption{Fundamental diagram for $\vm =5$ and $p=1/2$. Shown are the results
of the simulations and the mean-field theory.}
\label{fig:vmax5}
\end{figure}

In the site-oriented approach the state of the system is characterized by
the probability $P_n(t)$ to find at time $t$ a site in the state $n$, where
$n=0$ for an empty site and $n=1,2,\ldots,\vm$ for a site
occupied by a car with velocity $n$\ \footnote{Remember that we do not have
cars with velocity 0 due to the ordering 2-3-4-1.}.
Since in mean-field theory one neglects correlations between $n$ neighbouring
sites one can write down iteration equations for the $P_n$ in the
stationary state \cite{ssni}. These equations then allow a calculation of
the fundamental diagram. In Fig.\ \ref{fig:vmax5} results for the
``realistic'' value~\cite{ns} $\vm =5$ are compared with the simulations. The
mean-field flow is much too small showing the importance of correlations.

The results of mean-field theory can be improved systematically by the
so-called $n$-cluster approximation \cite{ss,ssni}. The $n$-cluster
method takes into account correlations between neighbouring sites and
reduces to mean-field theory for $n=1$. In order to get a closed set
of equations one uses conditional probabilities for the overlap of
neighbouring clusters \cite{ss,ssni}. Unfortunately, one then has to deal
with a system of $(\vm +1)^n$ nonlinear equations which in general cannot
be solved analytically.

Surprisingly, for $\vm =1$ the 2-cluster approximation already yields the
{\em exact} result. The probabilities $P(\sigma_1,\sigma_2)$ to find a
2-cluster in a state $(\sigma_1,\sigma_2)$ (where -- due to the ordering
2-3-4-1 -- $\sigma_j =0$ denotes an empty site and $\sigma_j =1$ a site
occupied by a car with velocity 1) are given by
\begin{eqnarray}
P(1,0)&=&P(0,1)={1-\sqrt{1-4qc(1-c)}\over 2q},\nonumber\\
P(0,0)&=&1-c-P(1,0),\qquad P(1,1)=c-P(1,0),
\end{eqnarray}
where $c=N/L$ is the density of cars and $q=1-p$. The corresponding flow is
just $f(c,p)=qP(1,0)$, i.e.
\begin{equation}
f(c,p)=\frac{1-\sqrt{1-4qc(1-c)}}{2}.
\label{fund1}
\end{equation}
Note that -- in contrast to what one finds using random-sequential dynamics --
parallel dynamics yield the well-known ``bunching'' observed in real
traffic, i.e.\ empty sites and cars attract each other ($P(0)P(1)=c(1-c)
\leq P(1,0)$). This means that there exists an attraction between two cars
separated by an empty site.\\
\begin{figure}
\vskip 60mm
\includegraphics{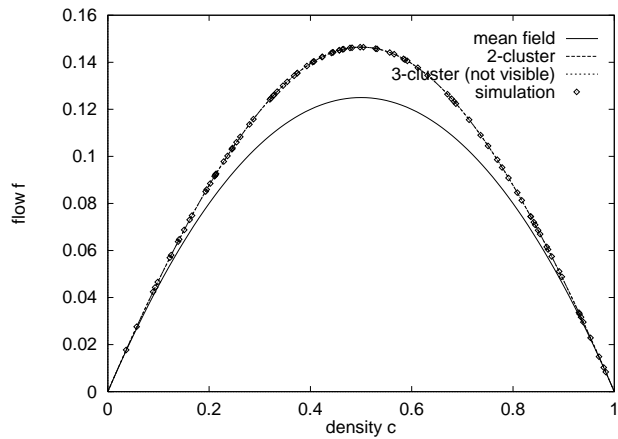}
\caption{Fundamental diagram for $\vm =1$ and $p=1/2$. Shown are the results
of the simulations and the 1- and 2-cluster approximations. The 2-cluster
result is exact.}
\label{fig:vmax1}
\end{figure}

\begin{figure}
\vskip 60mm
\includegraphics{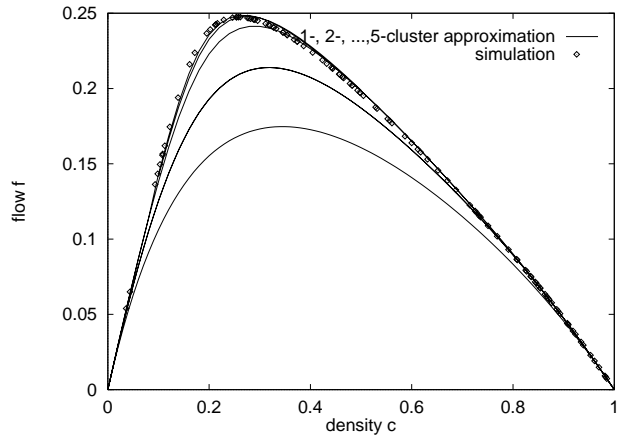}
\caption{Fundamental diagram for $\vm =2$ and $p=1/2$. Shown are the results
of the simulations and the $n$-cluster approximations for $n=1,2,\ldots 5$.}
\label{fig:vmax2}
\end{figure}

For higher velocities the $n$-cluster approximation does not become exact
reflecting the fact that long-range correlations exist for $\vm > 1$.
In Fig.\ \ref{fig:vmax2} we compare the results of the $n$-cluster
approximation for $n=1,\ldots,5$ with simulations. The $n$-cluster results
converge quite fast (the difference between the results for $n=4$ and $n=5$
is less than 1\%) and already the $n=5$ result is in very good agreement
with the simulations.

\section{Car-Oriented Approach}

In the car-oriented mean-field approach one introduces probabilities
$D_n(t)$ for finding at time $t$ exactly $n$ empty sites in front of a
vehicle.  In this way we already have taken into account correlations
between neighbouring sites. For $\vm =1$ the stationary solution of the
equations gouverning the time evolution of these probabilities can be
obtained quite easily \cite{caror}. One finds
\begin{eqnarray}
  D_0&=&\frac{2qc-1+\sqrt{1-4qc(1-c)}}{2qc},\nonumber\\
  D_n&=&\frac{D_0}{p}\left(\frac{p(1-D_0)}{D_0+p(1-D_0)}\right)^n
  \qquad\qquad (n\geq 1).
\end{eqnarray}
Using this result the flux can be calculated from
\begin{equation}
  f(c,p)=qc\sum_{n\geq 1}D_n.
\end{equation}
In this way one recovers the result (\ref{fund1}), i.e.\ in the
car-oriented approach already the mean-field result is exact.

\section{A Cellular Automaton Related to the Kasteleyn-Model}

In this Section we introduce a modified cellular automaton model in which
the cars have no maximum velocity. Starting from the model for $\vm =1$ as
described in Sect.\ 1 we modify the rules slightly. Suppose that we have
applied the rules 1-4 to a specific car. If the car does not move in step 4
nothing is changed. But if the car actually drives in step 4 we carry out
the car motion (by one site) and go back to step 2. This is repeated as
long as the car actually drives in step 4. This change now allows cars to
drive arbitrary distances (smaller than $d$ where $d$ is the distance to
the next car ahead), i.e.\ it corresponds to a maximum velocity $\vm
=\infty$.

It is interesting to note that this modified CA model is closely related
\cite{prizzi} to the so-called Kasteleyn model \cite{kast}.  This
model is a two-dimensional dimer model on a hexagonal lattice. By choosing the
activities in an appropriate way it is possible to map the dimer
configurations onto trajectories of a one-dimensional traffic problem
with periodic boundary conditions \cite{prizzi}.

Again the fundamental diagram of this modified cellular automaton can be
calculated exactly \cite{prizzi}. Already the mean-field result is exact
and yields
\begin{equation}
  f(c,p)=\frac{c(1-c)p}{1-(1-c)p} =\frac{c(1-c)\bar v}{1+c\bar v},
\label{ca_modif}
\end{equation}
where $\bar v = p/q$ is the mean velocity of free traffic.

\begin{figure}
\vskip 60mm
\includegraphics{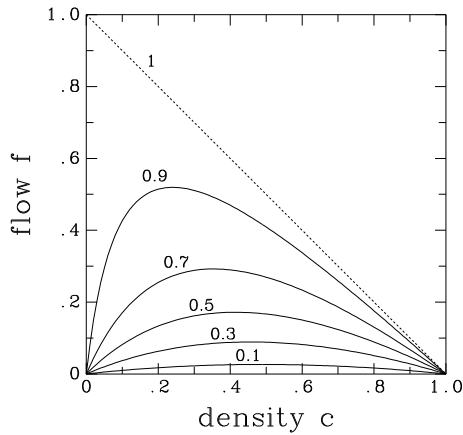}
\caption{Fundamental diagram for the modified CA for different values of
$p$.
\label{fig:prizzca}}
\end{figure}

\section*{Acknowledgments}

We would like to thank K.\ Nagel, N.\ Ito, V.B.\ Priezzhev and J.G.\ Brankov
for their collaboration on part of the results presented here.

\end{document}